\documentclass[5p,twocolumn]{elsarticle} %
\usepackage{verbatim}
\usepackage{lineno,hyperref}
\modulolinenumbers[5]
\usepackage{color}

\usepackage[]{subfigure}
\usepackage{placeins} 
\usepackage[english]{babel}  
\usepackage[utf8x]{inputenc}
\usepackage{graphicx}
\usepackage{booktabs}
\usepackage{tabularx}
\usepackage{caption}
\usepackage{amsmath}
\usepackage{amssymb}
\usepackage{multirow}

\begin{document}

\begin{frontmatter}

\title{On deformation of carbon nanotubes with TersoffCG: a case study}

\author[]{A. Pedrielli}

\begin{abstract}
Recently, TersoffCG, a coarse grain potential for graphene based on Tersoff potential, has been developed. In this work, we explore this potential, applying it to the case study of a single wall carbon nanotube. We performed a series of molecular dynamics simulations of longitudinal tension and compression on armchair carbon nanotubes, comparing two full atomistic models, described by means Tersoff and AIREBO potentials, and the coarse grained model described by means of TersoffCG. We followed each stage and mode of deformation, finding a good matching between the stress strain curves under tension independently from the used potential, with a small difference in the pre-fracture zone. Conversely, under compression the coarse grain model presents a buckling stress almost the double of the full atomistic models, and a more than double post-buckling stress. With the increase of the nanotube diameter, the capturing of the buckling modes is enhanced, however the stress overestimation remains. A decreasing of the three body angular term in the potential can be a rough way to recover the buckling stress, with small losses in the capturing of the post-buckling behavior. In spite of a good agreement under compression, the fracture behavior of the nanotube is strongly influenced, suggesting this modification only when no fractures are present. The findings reported in this work underlie the necessity of accurately evaluate the use of a coarse grain model when compressive loads are applied to the system during the simulation. 
\end{abstract}


\end{frontmatter}


\section{Introduction}

In recent years, a increasing interest has been devoted to 3D graphene structures \cite{Tylianakis2011,Wang2014} as means to deliver notable graphene mechanical properties to the macroscale. Among these structures, graphene nanofoams \cite{Wu2013, Qin2017, Pedrielli2018} and carbon nanotube networks \cite{Hall2008} have emerged as chemically stable, lightweight porous materials. Some efforts were done to pass from the study of general properties of these structures to their optimization as functional materials \cite{Xie2011, Pedrielli2017}. Bridging the analysis at the nanoscale and that at the macroscale, needs methods capable to deal with an high number of atoms, still capturing the main mechanical features of these materials. Recently, new potentials for graphene were developed with a coarse grain approach \cite{Cranford2009, Cranford2011, Ruiz2015}. Some of these potentials are also defined to be recursively extent at higher order of coarse graining \cite{Zhu2014}.
 However, each time we use a coarse grain model, we lose some information on the system with respect to the full atomistic description. In this work we evaluate the influence of using a recently developed a coarse grain potential for graphene \cite{Shang2017}, based on Tersoff potential, named TersoffCG, on the case study of a single wall carbon nanotube. We computed the stress strain curves under tension and compression along longitudinal direction, comparing the coarse grain model and two full atomistic models described by means of Tersoff \cite{Tersoff1988, Tersoff1988A} and AIREBO \cite{Stuart2000} potential. Following step by step the tension and the compression of the nanotube we also evaluated the buckling modes, and their possible suppression due to the use of the coarse grain model. The findings of this work show how the coarse grain model accurately works under tension while a loss of accuracy is found under compression. 
 \begin{table*}[htbp]
\centering
\begin{tabular}{lccccccccccccccc}
\toprule
 m & $\gamma$ & $\lambda_3$ & c & d & $\cos \theta_0$ & n & $\beta$ & $\lambda_2$ & B & R & D & $\lambda_1$ & A \\
 \midrule
 3 & 1 & 0 & 38049 & 4.3484 & -
0.57058 & 0.72751 & 1.572x10$^-7$ & 1.10595 & 1386.8 & 4.1 & 0.6 & 1.74395 & 5574.4\\
 \bottomrule
\end{tabular}
\caption{Parameters of TersoffCG potential for carbon, as developed in by Shang et al. \cite{Shang2017}}.
\label{tab:Parameters}
\end{table*}

\section{Computational model }

The case study we use here is a $5$~nm $(20,20)$ armchair nanotube. For the coarse grain model we use instead a $(10,10)$ armchair nanotube with doubled bond length. In this way, as asked by the coarse grain model, we have, at the first order, a coarse grain atom in place of four atoms in the full atomistic model. At the same time the coarse grain atoms have a mass four times that of a carbon atom. The unit cell sides were fixed to $6$~nm along the directions perpendicular to the nanotube axis. We used for TersoffCG\cite{Shang2017}, Tersoff \cite{Tersoff1988, Tersoff1988A} and AIREBO \cite{Stuart2000} potentials the standard parametrizations, without taking into account typical internal cutoff for the near fracture regimes \cite{Shenderova2000}. This, anyway have no influence on the results presented here. Molecular dynamics simulations were performed within LAMMPS code \cite{Plimpton1995}.
The parameters used for TersoffCG \cite{Shang2017} are reported in Tab. \ref{tab:Parameters}.

We imposed periodic boundary conditions on the longitudinal direction and applied the deformation in the same direction. 
All the samples were initially fully relaxed, then equilibrated to zero pressure and the target temperature $1$ or $300$~K. All the mechanical tests, were performed with successive $0.01$ deformation steps followed by $2$~ps isothermal ensemble (NVT) equilibration by means of Nosé–Hoover thermostat. The stress was computed and averaged during a NVT run of $1$~ps. In all the simulations the equations of motion were solved with the velocity-Verlet integration method using a time step of $1$~fs.

The engineering strain parallel to the direction of deformation is defined as
\begin{equation}
\varepsilon = \frac{L-L_0}{L} = \frac{\Delta L}{L}
\end{equation} 
where $L_0$ and $L$ are the starting and current length of the sample in the direction of loading. 
To determine the stress, the pressure stress tensor components in response to the
external deformation are computed as \citep{Thompson2009}
\begin{equation}\label{pressure}
P_{ij} = \frac{\sum_k^N{m_k v_{k_i} v_{k_j}}}{V}+ \frac{\sum_k^N{r_{k_i} f_{k_j}}}{V}
\end{equation} 
where $i$ and $j$ label the coordinates $x$, $y$, $z$; $k$ runs over the
atoms; $m_k$ and $v_k$ are the mass and velocity of $k$-th atom; $r_{k_i}$ is the position of $k$-th atom; $f_{k_j}$ is the $j$-th component of the total force on the $k$-th atom due to the other atoms; and, finally, $V$ is the volume of the simulation box. We note that the stress presented along this paper is referred to the simulation cell sectional area of $36$ nm$^2$. 
The pressure in Eq. \ref{pressure} includes both kinetic energy (temperature) and virial term.

In the same way a $10$~nm $(20,20)$ and $5$~nm and $10$~nm $(40,40)$ nanotubes were prepared and tested.

\section {Mechanical tests and buckling modes}

\begin{figure}[htbp]
\centering
{\includegraphics[width=0.5\textwidth]{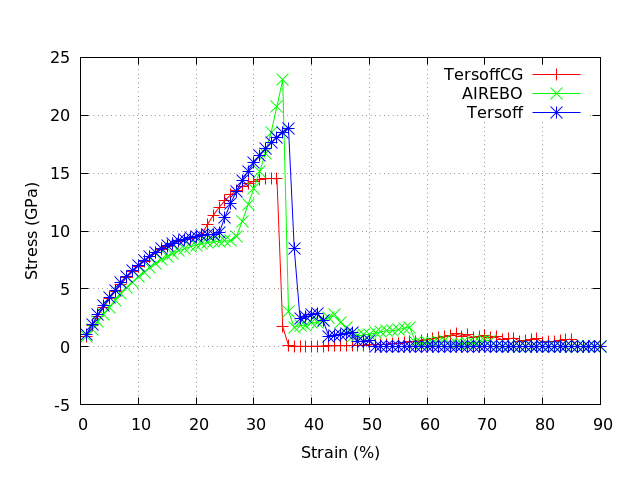}}
\caption{Stress strain curves under tension of the considered $5$~nm $(20,20)$ armchair nanotube for full atomistic and coarse grain potentials. The used temperature was $1$~K. We found an overall matching of the stress strain curve for the three potentials, with a similar fracture strain. }
\label{fig:Tens1K}
\end{figure}

We report in Fig. \ref{fig:Tens1K} the stress strain curves under tension of the considered $5$~nm $(20,20)$ armchair nanotube for full atomistic and coarse grain potentials, at $1$~K temperature. We found an overall matching of the stress strain curve for the three potentials, with a similar fracture strain.

\begin{figure}[htbp]
\centering
{\includegraphics[width=0.5\textwidth]{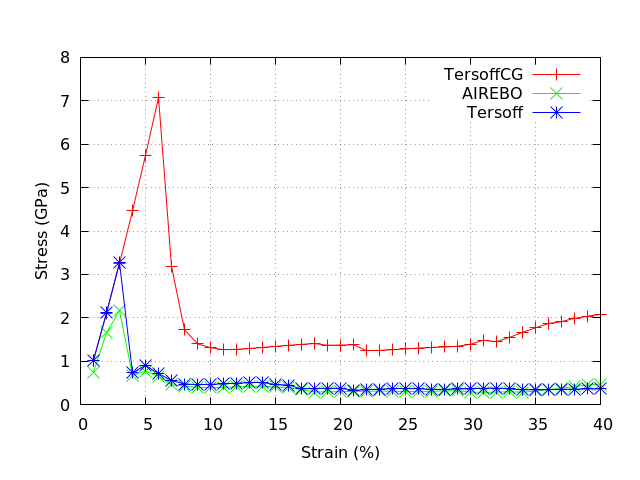}}
\caption{Stress strain curves under compression of the considered $5$~nm $(20,20)$ armchair nanotube for full atomistic and coarse grain potentials. The used temperature was $1$~K. The stress strain curves under compression are very sensitive with respect to the use of coarse grain model. Indeed apart form the slope of the elastic part, i.e. the Young modulus, the features are different. Fracture strain and fracture stress are almost the double with respect to those obtained for full atomistic potentials, the post buckling stress is instead more than double. }
\label{fig:Comp1K}
\end{figure}

Regarding the compressive case, we report in Fig. \ref{fig:Comp1K} the stress strain curves of the considered $5$~nm $(20,20)$ armchair nanotube for full atomistic and coarse grain potentials, at $1$~K temperature. The stress strain curves under compression are very sensitive with respect to the use of coarse grain model. Indeed, apart form the slope of the elastic part, i.e. the Young modulus, the features are different. Fracture strain and fracture stress are almost the double with respect to those obtained for full atomistic potentials, the post-buckling strain is instead more than double. The difference between the two full atomistic potentials is instead limited to the value of buckling stress presenting the same buckling strain and post-buckling stress strain curve.

\begin{figure}[htbp]
\centering
{\includegraphics[width=0.5\textwidth]{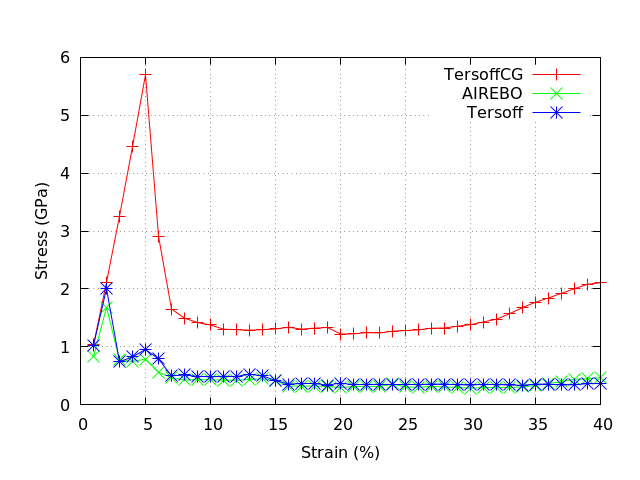}}
\caption{Stress strain curves under compression of the considered $5$~nm $(20,20)$ armchair nanotube for full atomistic and coarse grain potentials. The used temperature was $300$~K. The stress strain curves under compression are very sensitive with respect to the use of coarse grain model. Indeed apart form the slope of the elastic part, i.e. the Young modulus, the features are different. Fracture strain and fracture stress and the post buckling stress are almost three times those obtained for full atomistic potentials. }
\label{fig:Comp300K}
\end{figure}

In order to evaluate the influence of the temperature on the different behavior of the coarse grain model with respect to the full atomistic ones and that in buckling stress between the two atomistic models, we performed the same compressive tests at $300$~K temperature (Fig. \ref{fig:Comp300K}).  Fracture strain and fracture stress and the post buckling with the coarse grain model are almost three times those obtained for full atomistic potentials. The effect of the temperature is to level out the buckling stress for atomistic models, whereas the coarse grain model is essentially uninfluenced
(Fig. \ref{fig:Comp300K}).

\begin{figure}[htbp]
\centering
{\includegraphics[width=0.4\textwidth]{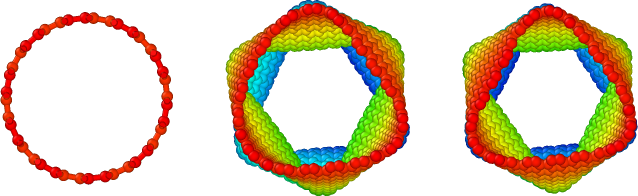}\\
 \vspace{1cm}
\includegraphics[width=0.4\textwidth]{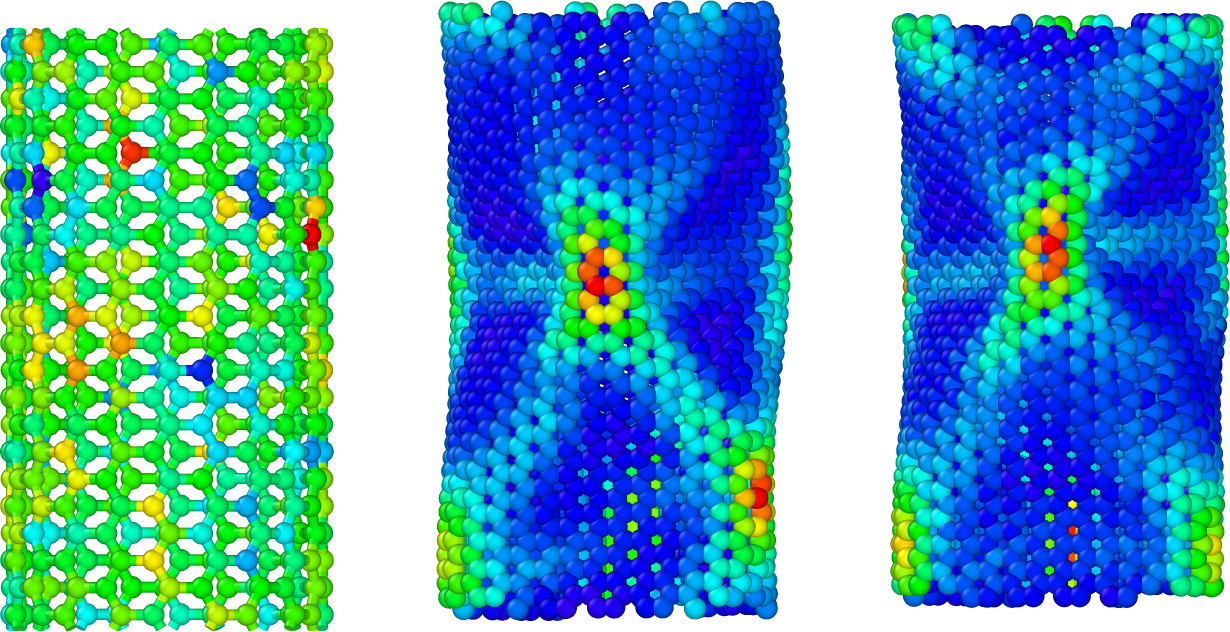}}
\caption{Section and transverse views of the three model of nanotube for $0.04$ strain. From the left, TersoffCG, Tersoff and AIREBO samples. Color code is arbitrary in upper panel, potential energy in the lower one.}
\label{fig:Comp40}
\end{figure}

\begin{figure}[htbp]
\centering
{\includegraphics[width=0.4\textwidth]{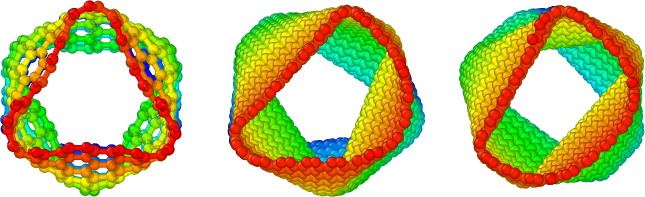}\\
 \vspace{1cm}
\includegraphics[width=0.4\textwidth]{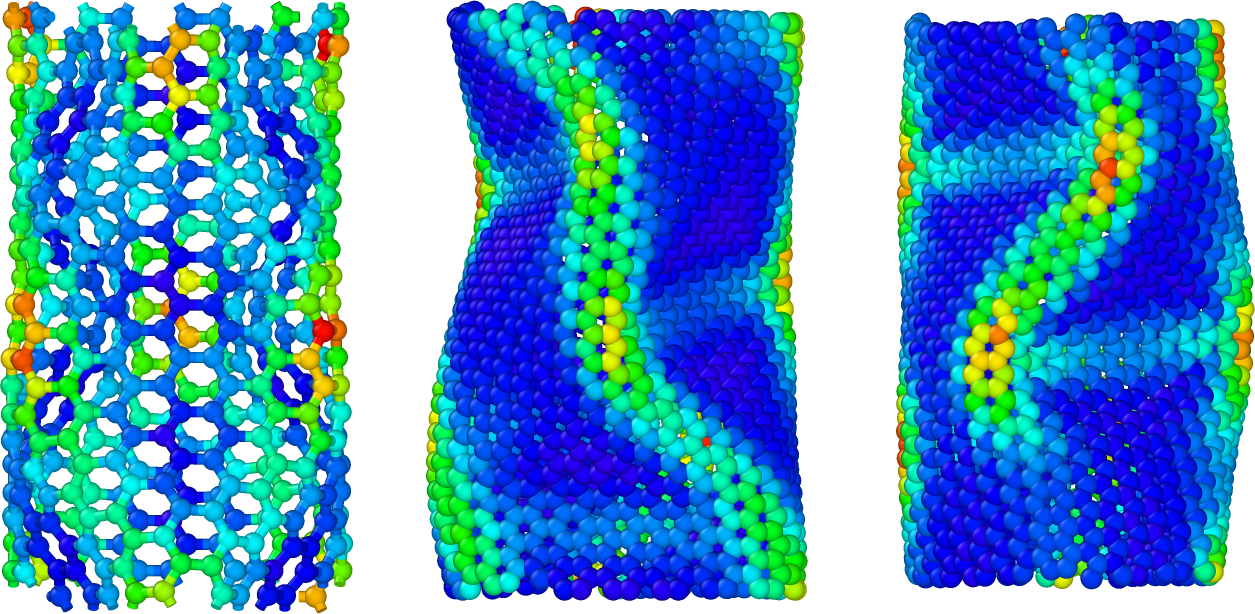}}
\caption{Section and transverse views of the three model of nanotube for $0.07$ strain. From the left, TersoffCG, Tersoff and AIREBO samples. Color code is arbitrary in upper panel, potential energy in the lower one.}
\label{fig:Comp70}
\end{figure}

\begin{figure}[htbp]
\centering
{\includegraphics[width=0.4\textwidth]{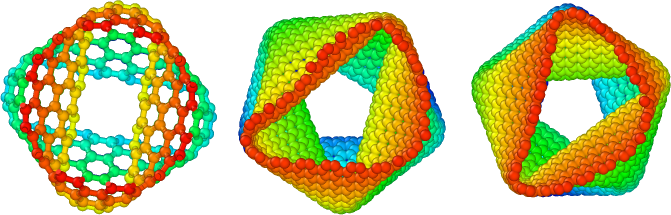}\\
 \vspace{1cm}
\includegraphics[width=0.4\textwidth]{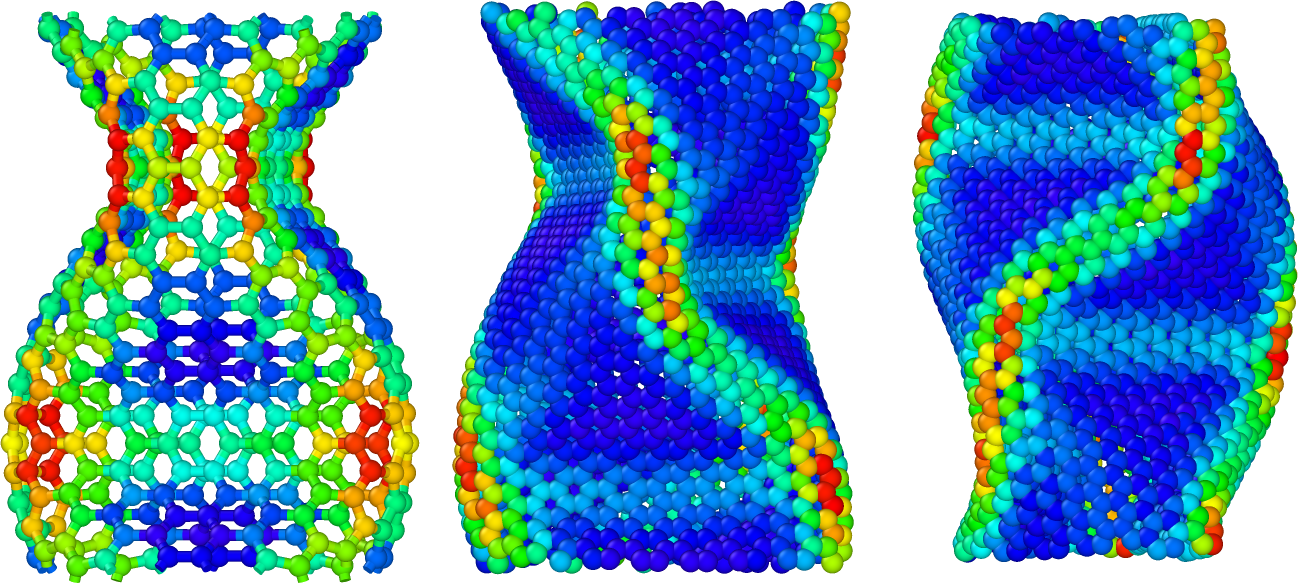}}
\caption{Section and transverse views of the three model of nanotube for $0.12$ strain. From the left, TersoffCG, Tersoff and AIREBO samples. Color code is arbitrary in upper panel, potential energy in the lower one.}
\label{fig:Comp120}
\end{figure}

\begin{figure}[htbp]
\centering
{\includegraphics[width=0.4\textwidth]{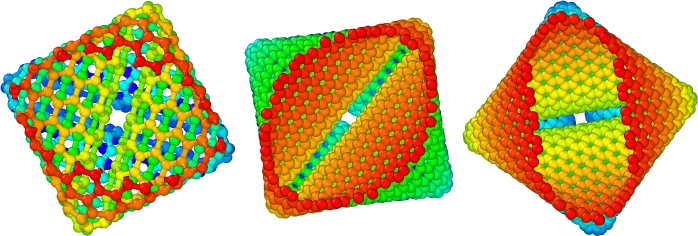}\\
 \vspace{1cm}
\includegraphics[width=0.4\textwidth]{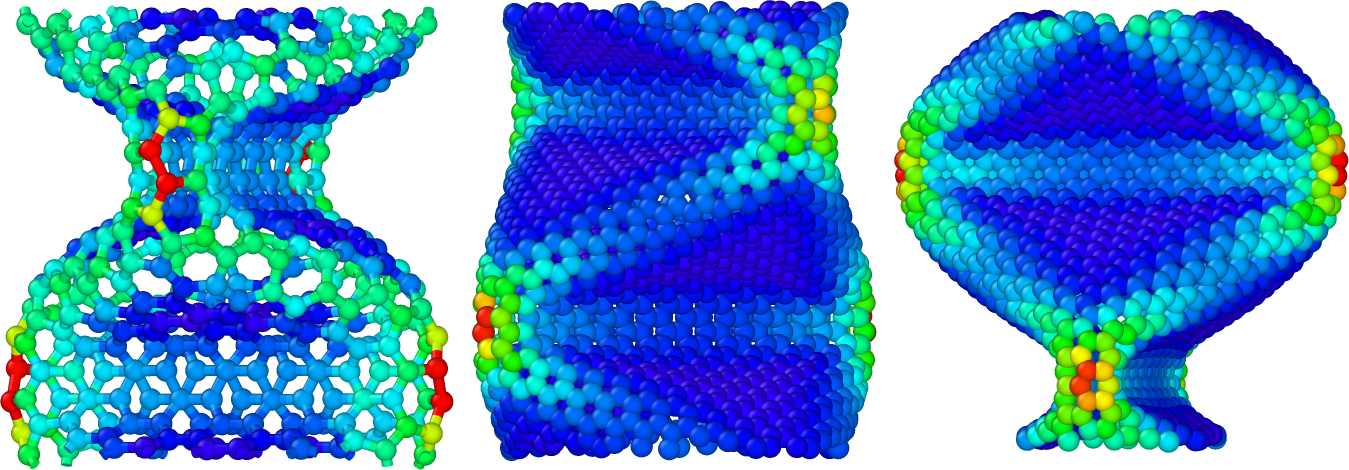}}
\caption{Section and transverse views of the three model of nanotube for $0.35$ strain. From the left, TersoffCG, Tersoff and AIREBO samples. Color code is arbitrary in upper panel, potential energy in the lower one.}
\label{fig:Comp350}
\end{figure}

We report in Fig. \ref{fig:Comp40}, \ref{fig:Comp70}, \ref{fig:Comp120}, \ref{fig:Comp350} the nanotube samples at $4$, $7$, $12$ and $35$\% compressive strain. In each figure we show in the upper panel longitudinal view of the nanotube cell, in the lower panel a front view of the nanotube. These particular strain values where chosen to show the difference in the buckling mode, and the strain at which they appear. We have seen from the stress strain curves in Fig. \ref{fig:Comp1K} that the first buckling mode is obtained for full atomistic models between $2$ and $5$\% strain, we report in Fig.  \ref{fig:Comp40} the three nanotubes at $4$\% strain showing the first buckling mode of the full atomistic models, whereas the coarse grain model is still in pre-buckling.

A second stage is that presented at $7$\% strain, reported in Fig. \ref{fig:Comp70}, for which a second buckling mode is obtained for full atomistic models, corresponding to the second small peaks in the stress strain curves reported in Fig. \ref{fig:Comp1K}. For the coarse grain model, at $7$\% strain we have the appearance, with a certain delay, of the first mode, noted for atomistic models in at $4$\% strain. 

In the third stage, presented in Fig. \ref{fig:Comp120} at $12$\% strain, the buckling mode of the atomistic models is essentially the same of $7$\% strain, whereas the coarse grain model directly pass to the last buckling mode, that it will keep at the subsequent strain increases. 

In the last stage, presented in Fig. \ref{fig:Comp120} at $35$\% strain, the buckling mode of the atomistic model and that of the coarse grain model are the same. 

The coarse grain models presents only two of the three buckling modes presented by both atomistic models, the first mode and the third. The buckling mode presented by atomistic samples in Fig. \ref{fig:Comp70} and \ref{fig:Comp120} is eliminated using the coarse grain potential.

\begin{figure}[htbp]
\centering
{\includegraphics[width=0.5\textwidth]{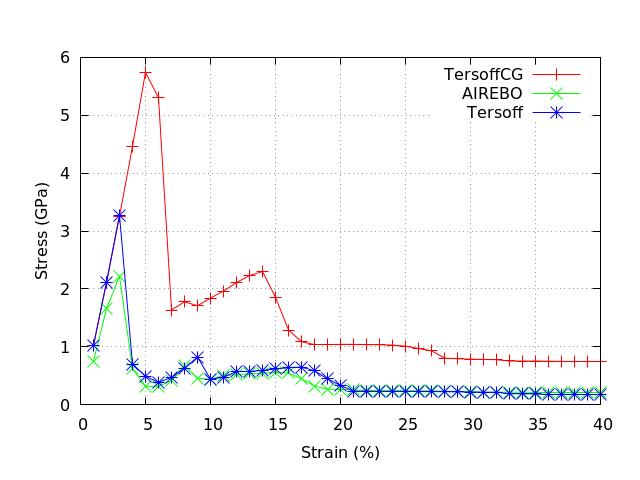}}
\caption{Stress strain curves under compression of the considered $10$~nm $(20,20)$ armchair nanotube for full atomistic and coarse grain potentials. The used temperature was $1$~K. The stress strain curves under compression are very sensitive with respect to the use of coarse grain model. The main features are present also in the coarse grain model, however all the peaks corresponding to subsequent buckling instabilities are found at higher stress and higher strain than that for the full atomistic models.}
\label{fig:CompLong}
\end{figure}

The same tests were performed on a $10$~nm $(20,20)$ carbon nanotube and its coarse grain analogous, with similar findings. In particular, as shown in Fig. \ref{fig:CompLong}, also in this case the stress strain curves under compression are very sensitive with respect to the use of coarse grain model. The main features are present also in the coarse grain model, however all the peaks corresponding to subsequent buckling instabilities are found at higher stress and higher strain than that for the full atomistic models. 

\begin{figure}[htbp]
\centering
{\includegraphics[width=0.5\textwidth]{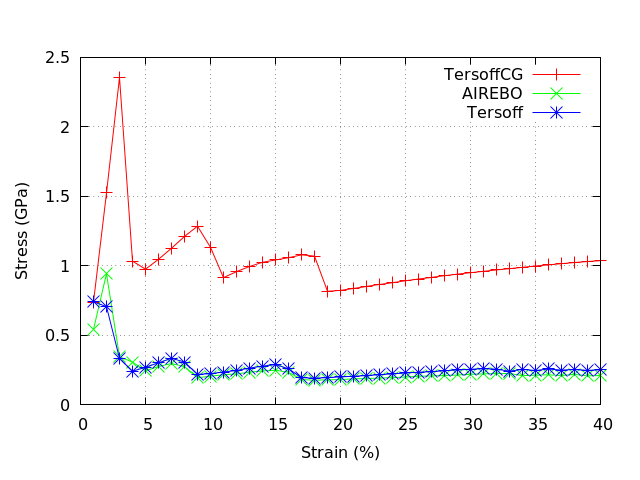}}
\caption{Stress strain curves under compression of the considered $5$~nm $(40,40)$ armchair nanotube for full atomistic and coarse grain potentials. The used temperature was $1$~K. The stress strain curves under compression are very sensitive with respect to the use of coarse grain model. As for the double length nanotube of which stress strain curves are reported in Fig. \ref{fig:CompLong} the main features are present also in the coarse grain model, in this case the peaks corresponding to subsequent buckling instabilities are found at higher stress and almost the same strain than that for the full atomistic models.}
\label{fig:CompLarge}
\end{figure}

A further test was done on a $5$~nm $(40,40)$ carbon nanotube (Fig. \ref{fig:CompLarge}). As for the double length nanotube of which stress strain curves are reported in Fig. \ref{fig:CompLong} the main features are present also in the coarse grain model, in this case the peaks corresponding to subsequent buckling instabilities are found at higher stress and almost the same strain than that for the full atomistic models.

\FloatBarrier

\section{A rough modification of TersoffCG for compressive regime}

\begin{figure}[htbp]
\centering
{\includegraphics[width=0.5\textwidth]{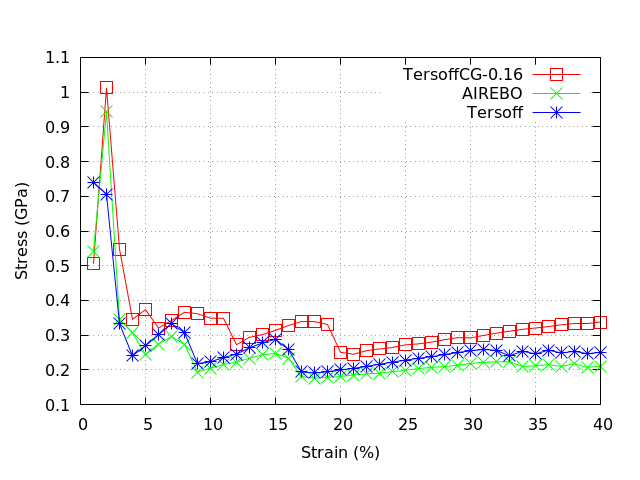}}
\caption{Stress strain curves under compression at $1$~K temperature of the considered $5$~nm $(40,40)$ armchair nanotube for full atomistic and modified TersoffCG potentials with a value of $\gamma$ parameter equal to $0.16$. The general trend reported for TersoffCG, reported in Fig. \ref{fig:CompLarge} is maintained but the agreement of the stress values with the full atomistic simulations is strongly improved.}
\label{fig:Modify}
\end{figure}

\begin{figure}[htbp]
\centering
{\includegraphics[width=0.5\textwidth]{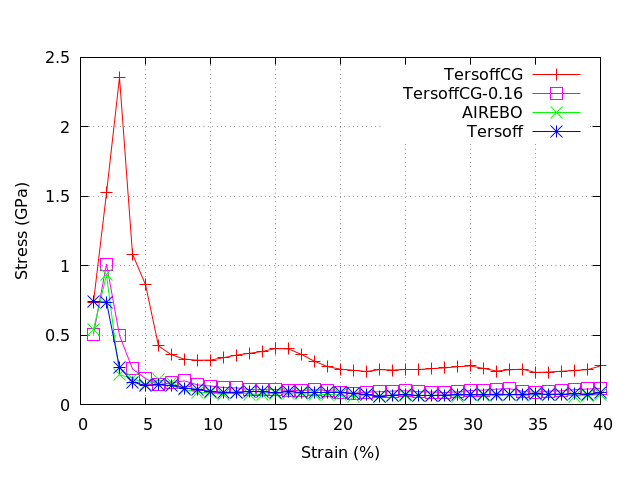}}
\caption{Stress strain curves under compression at $1$~K temperature of the considered $10$~nm $(40,40)$ armchair nanotube for full atomistic and modified TersoffCG potentials with a value of $\gamma$ parameter equal to $0.16$. The general trend reported for TersoffCG, is maintained but the agreement of the stress values with the full atomistic simulations is strongly improved.}
\label{fig:Modify1}
\end{figure}

For the $5$~nm $(40,40)$ nanotube, the main peaks in the stress strain curves, are found in good agreement with those obtained with full atomistic models. In this case, to increase the nanotube diameter enhances the agreement the the capturing of the peaks positions in the stress strain curves in Fig.  \ref{fig:CompLarge} without improving the accuracy on the stress values.
Here we indicate a rough modification to TersoffCG potential that can improve the matching of the stress values under compression. As each modification to a complex potentials, it should be verified accurately with respect to the aim of the simulations. The parameter of TersoffCG that we modified is $\gamma$, the multiplicative factor on the angular term (Tab. \ref{tab:Parameters}).
In Fig. \ref{fig:Modify} we report the stress strain curve under compression, obtained for the $5$~nm $(40,40)$ nanotube at $1$~K a value of $\gamma$ parameter equal to $0.16$. This modification strongly affects the near- and post- fracture regime. Anyway, under compression the change results to be effective. 
A further test was performed on a $10$~nm $(40,40)$ nanotube, with good agreement between the modified TersoffCG end the full atomistic potentials. The stress strain curves are reported in Fig. \ref{fig:Modify1}.

\FloatBarrier

\section{Conclusions}

In this work, we applied the TersoffCG potential to the case study of a single wall carbon nanotube. The general performance of the potential was tested under tension and compression comparing the results with those obtained using two full atomistic models. Under tension the results for the various potentials are overall similar, and a easy application of the coarse grain model could be safe. Conversely, under compression the only features of the stress strain curves that resemble the atomistic model are the Young modulus and the general trend. However buckling strain, buckling stress and post-buckling stress are overestimates by a factor two. 

With regards to the buckling modes, one of those presented by the $5$~nm $(20,20)$ nanotube in the atomistic models is suppressed. Furthermore, the strain at which the modes appear is different when we use the coarse grain model. 
With the increasing of the nanotube length ($5$~nm $(20,20)$ nanotube), the main buckling modes are captured also by the coarse grain model. Increasing instead the nanotube diameter ($5$~nm $(40,40)$ nanotube) the critical strain at which the buckling modes appear better matches that of full atomistic models.

The overestimation of the buckling stress can be roughly cured modifying the parameter $\gamma$ in TersoffCG potential. This is effective under compression and we successfully reproduce the stress strain curves for a $5$~nm $(40,40)$. However, the choice of the value of $\gamma$ could be system dependent. Furthermore, under tensile stress, in the near- and post- fracture regime the modification produces unphysical results. The modification of $\gamma$ is then limited to specific compressive problems.

From this case study we have seen that the tension case can be easily treated with TersoffCG whereas, under compression, the presence of instabilities and the high deformation of the nanotube structure at the wrinkles suggest more attention in the use of coarse grain potentials. Finally, the compressive test of carbon nanotube could be taken as a useful test for coarse grain graphene potentials.

\FloatBarrier


\bibliography{Bibliography.bib}

\end{document}